\documentclass[12pt,preprint]{aastex}

\shorttitle{MAGNETIC HELICITY AND MAJOR SOLAR FLARES}
\shortauthors{PARK et al.}
\begin{document}

\title{THE VARIATION OF RELATIVE MAGNETIC HELICITY AROUND MAJOR FLARES}

\author{SUNG-HONG PARK,\altaffilmark{1,2} JEONGWOO LEE,\altaffilmark{1} GWANG-SON CHOE,\altaffilmark{4,5} JONGCHUL CHAE,\altaffilmark{3} HYEWON JEONG,\altaffilmark{3} GUO YANG,\altaffilmark{1,2} JU JING,\altaffilmark{1,2} AND HAIMIN WANG\altaffilmark{1,2}}

\altaffiltext{1}{Center for Solar-Terrestrial Research, New Jersey Institute of Technology, 323 Martin Luther King Boulevard, 101 Tiernan Hall, Newark, NJ 07102; sp295@njit.edu.}
\altaffiltext{2}{Big Bear Solar Observatory, 40386 North Shore Lane, Big Bear City, CA 92314.}
\altaffiltext{3}{Astronomy Program and FPRD, Department of Physics and Astronomy, Seoul National University, Seoul 151-742, Korea.}
\altaffiltext{4}{Department of Astronomy and Space Science, Kyung Hee University, Yongin 449-701, Korea.}
\altaffiltext{5}{Princeton Plasma Physics Laboratory, Princeton, NJ 08543-0451.}

\begin{abstract}

We have investigated the variation of magnetic helicity over a span
of several days around the times of 11 X-class flares which
occurred in seven active regions (NOAA 9672, 10030, 10314, 10486,
10564, 10696, and 10720) using the magnetograms taken by the
Michelson Doppler Imager (MDI) on board the $Solar$ $and$ $Heliospheric$
$Observatory$ ($SOHO$). As a major result we found that each of these
major flares was preceded by a significant helicity accumulation,
(1.8--16)$\times$10$^{42}$ Mx$^2$ over a long period (0.5 to a few
days). Another finding is that the helicity accumulates at a nearly
constant rate, (4.5--48)$\times$10$^{40}$ Mx$^2$ hr$^{-1}$, and then
becomes nearly constant before the flares. This led us to
distinguish the helicity variation into two phases: a phase of
monotonically increasing helicity and the following phase of relatively
constant helicity. As expected, the amount of helicity accumulated
shows a modest correlation with time-integrated soft X-ray flux
during flares. However, the average helicity change rate in the
first phase shows even stronger correlation with the time-integrated
soft X-ray flux. We discuss the physical implications of this result
and the possibility that this characteristic helicity variation
pattern can be used as an early warning sign for solar eruptions.

\end{abstract}


\section{INTRODUCTION}

Magnetic helicity is a measure of twists, kinks, and inter-linkages
of magnetic field lines (Berger \& Field 1984) and has been an
important parameter in solar dynamo theories (Parker 1955). While
the source of the magnetic helicity lies below the surface of the
Sun, it was recently recognized as a useful parameter in describing
solar features observed above surface such as spiral patterns of
sunspot fibrils, helical patterns in filaments and coronal mass
ejections (CMEs; for a review, see Brown et al. 1999).
Naturally magnetic helicity studies have been directed to the
energy buildup and instability leading to eruptions and CMEs (e.g.,
Rust 2001; Kusano et al. 2004; Phillips et al. 2005). More recently,
several studies were carried out to relate the change of magnetic
helicity to the problem of impending or triggering solar flares.
Moon et al. (2002a) studied the magnetic helicity change around
major flares to find its rapid helicity change before flares, and
concluded that a sudden helicity injection may trigger flares. Moon
et al. (2002b) applied the same approach to seven homologous flares
in the active region, NOAA 8100, over a period of 6.5 hr to find
a good correlation between the amount of incremental helicity and
the soft X-ray flux during each homologous flare. The results from
both studies thus point to the idea that the helicity change
occurring over short timescales (around a half hour) can be a
significant factor in triggering flares. Kusano et al. (2003)
proposed annihilation of magnetic helicity as a triggering mechanism
for solar flares. Numerical simulations were carried, which show
that, if the helicity is sharply reversed within a magnetic arcade,
reconnection quickly grows in the helicity inversion layer, driving
explosive dynamics. Yokoyama et al. (2003) studied flare activities
in active region NOAA 8100 to find that most of the flare events
occurred about half a day after the helicity injection rate changed
its sign, and the positions of H$\alpha$ emission in flares well
correspond to the helicity inversion lines in space.  Sakurai \&
Hagino (2003) studied two active regions appeared in 2001 (NOAA 9415
and 9661), both of which have produced X-class flares. Their finding
was, on the contrary, that the magnetic helicity integrated over the
regions evolved slowly and did not show abrupt changes at the time
of the flares, although the distributions of magnetic helicity
changed significantly over a few days in the regions.

In this paper, we study long term (a few days) variations of the
magnetic helicity around major X-class flares. While some of the
above studies suggest short-term helicity change as an important
topic for flare triggering, Hartkorn \& Wang (2004) found that the
rapid helicity change at the time of a flare can occur as an
artifact under the influence of flare emission on the spectral line
adopted in MDI measurements. This means that a short term variation
during strong flares can hardly be measured with enough accuracy.
However we can study a long-term variation of the magnetic helicity
with active region magnetograms gathered over many days excluding
the times of flares. Our research is therefore focused on a possible
characteristic helicity evolution pattern that is associated with
flare impending mechanisms.

\section{CALCULATION OF MAGNETIC HELICITY}

By magnetic helicity, we refer to the {\it relative} magnetic
helicity in the rest of this paper, i.e., the helicity relative to
that of the potential field state. With the time-dependent
measurement of longitudinal magnetic fields in the photosphere, we
can only approximately determine the change rate of the relative
magnetic helicity (Demoulin \& Berger 2003). We further use a
simplified expression for the helicity change rate (Chae 2001) given
by

\begin{equation}
\left(\frac{dH_r}{dt}\right)_{LCT} = -\int_s 2({\bf{A}}_p \cdot
{\bf{v}}_{LCT})B_n dS
\end{equation}
where $B_n$ is the normal component of the magnetic field; ${\bf
A}_p$ is the vector potential of the potential field; ${\bf
v}_{LCT}$ represents the apparent horizontal motion of field lines;
$dS$ is the surface integral element and the integration is over the
entire area of the target active region. Although this expression
does not explicitly include the helicity change by the vertical
motion of field lines (see Kusano et al. 2002), D\'emoulin \& Berger
(2003) pointed out that it actually accommodates both the vertical
and horizontal motions of flux tubes as far as no flux tube newly
emerges from or totally submerges into the surface. As we will show
later in presenting the result of helicity calculation, this
requirement is not always met.

We determine the quantities in the above equation using full disk
MDI (Scherrer et al. 1995) magnetograms following the procedure described in
Chae \& Jeong (2005). First, we approximately determine $B_n$ from
the line-of-sight magnetic field $B_l$ in the MDI magnetograms,
simply considering the projection effect, i.e., $B_l= B_n \cos\psi$
where $\psi$ is the heliocentric angle of the point of interest,
assuming that the magnetic field on the solar photosphere is normal
to the solar surface. Second, ${\bf A}_p$ is calculated from $B_n$
by using the Fast Fourier Transform method as usual. The extent of
the spatial domain of the Fourier transform is taken about twice
wider than the active region area in order to minimize the artifacts
arising from the periodic boundary condition in the fast Fourier
transform (Alissandrakis 1981). Third, ${\bf v}_{LCT}$ is
calculated using the local correlation tracking (LCT) technique
(November \& Simon 1988). For local correlation tracking, we align
all magnetograms in each event to the first image of the data set
after correcting the differential rotation. We set the FWHM of the
apodizing window function to 10{\arcsec} and the time interval
between two frames to 60 minutes, and performed LCT for all pixels with
an absolute flux density greater than 5 G. Only the pixels with
cross correlation above 0.9 are considered.

In selecting data we found that use of 1 minute cadence full-disk
MDI (Scherrer et al. 1995) magnetograms is adequate for our purpose of
investigating the long-term helicity evolution. However, there are
occasionally found data gaps in the 60 minute cadence data set in
which case we supplement the data gaps with 96 minute MDI
magnetograms. The time interval of the supplemented data set is
therefore not longer than 96 minute. To reduce the effect of the
geometrical projection, we selected the active regions lying within
60\% of the solar radius from the apparent disk center. Note that
we use only full disk MDI magnetograms that have 2{\arcsec}$\times$2{\arcsec}
pixel size. Therefore the LCT velocities calculated here may have been
systematically underestimated compared with the LCT velocities calculated
with the higher resolution (0.6{\arcsec}$\times$0.6{\arcsec}) MDI data (Longcope et al. 2007).

After the helicity change rate is determined as a function of time,
we integrate it with respect to time to determine the amount of
helicity accumulation:

\begin{equation}
\Delta H = \int_{t_0} ^{t} \left(\frac{dH_r}{dt}\right) dt
\end{equation}
where $t_0$ and $t$ are  the start and end time of the helicity
accumulation, respectively. If $t_0$ is a time when the magnetic
field is in the potential state, $\Delta H$ is simply the helicity,
$H(t)$, at time $t$. However, there is no guarantee that we can
observe, by chance, an active region in the potential energy state.
We therefore set $t_0$ as the earliest time without significant
helicity accumulation at which the average value of the helicity
change rate over four hours is less than our nominal threshold in
helicity change rate, 1$\times$10$^{40}$ Mx$^2$ hr$^{-1}$. If that
time cannot be determined, we define $t_0$ as the time when the data
set starts or when the previously accumulated helicity is released
by a flare. The exact time of $t_0$ here is unimportant because it
is only a trial value. After determining $H(t)$, we redefine $t_0$
as the time when the resulting helicity starts to increase from a
nearly constant value.

\section{Magnetic Helicity Variation}

We present the helicity variation calculated for seven active
regions in Figures 1 and 2. In both figures we plot the magnetic
helicity accumulation together with the GOES soft X-ray light curve
and magnetic flux as functions of time. The soft X-ray light curve is
shown to indicate the flare times and the magnetic flux is shown to
check the above-mentioned requirement for the approximation made in
Equation (1). Note that the fluxes shown in this paper are total
unsigned magnetic flux, i.e., sum of the absolute amounts of positive
and negative fluxes, because net magnetic flux may show little
change despite significant flux change in each polarity.

For the events shown in Figure 1 we can see that the helicity
accumulates at a monotonic rate of change about 0.5--2 days before the
flare onset, and then becomes almost constant before the flares. For
convenience, we distinguish the magnetic helicity variation in two
stages: a phase of monotonically increasing helicity (phase I) and the
following phase of relatively constant helicity (phase II). This
pattern is obvious for the four flares (2001 October 25, 2004 November 7, and
2005 January 16 and 17). For the 2005 January 15 event, the
helicity increased up to 22:00 UT on January 14 and then decreased
afterward. In this case, we do not consider that the flare occurred
in phase II. It is then noted that these flares took place
after a significant amount, $\sim$(1.8--11)$\times$10$^{42}$
Mx$^2$, of helicity accumulation.

In Figure 2, we show the result for the other four active regions.
Like the events in Figure 1, these flares also occurred after a
significant helicity accumulation, $\sim$(1.9--16)$\times$10$^{42}$
Mx$^2$. However, they occurred in the middle of the continuous
helicity accumulation, unlike those events shown in Figure 1. In
other word, the flares in Figure 2 occurred in phase I, while those
in Figure 1 occurred in phase II. One common trend is, however, that
all the events are apparently associated with a considerable amount
of helicity build-up before the flares, whether they occurred in
phase I or in phase II. In case where flares occur in phase II, it may
imply that solar active regions can wait for major flares after the
helicity accumulated to some limiting amount. This is seemingly
contrary to the general belief that a flare occurs as soon as the
system reaches some threshold. An active region may evolve to a
certain stage where the helicity no longer increases, and the system
waits until it unleashes the stored energy by producing flares due
to certain mechanism of triggering.

Since we claimed that these large flares are always preceded by
significant accumulation of helicity, as a reference we check
the corresponding helicity variation in non-flaring times. We show
such data in Figure 3. For all active regions under investigation,
the amount of helicity change during nonflaring periods (Fig. 3) is
much less than that around the major flare time. This convinces us
that the above monotonically increasing helicity before major flares is a
process associated with the flares and is not occurring in nonflare
times. Another point to note in Figure 3 is that not only the
helicity but also the total unsigned magnetic flux changes much less
during the nonflaring time compared with the period before major
flares. This implies that the increase of the magnetic helicity
before major flares is, in part, related to the simultaneous
increase of total unsigned magnetic flux.

It is also worthwhile to mention how the characteristic pattern of
the helicity variation found here will depend on the sign of
helicity. In our result obtained for seven active regions, similar
amounts of both of positive and negative helicity were accumulated
continuously and simultaneously during the whole time. It is
therefore unlikely that counting the helicity in one and the other
polarity separately yields a significantly different conclusion. On
the other hand, some studies suggested that the sign-reversal of the
helicity injection rate is important for flare activity and we need
to compared them with the present result. Kusano et al. (2003)
emphasized spatially sharp reversal of helicity sign triggers
magnetic reconnection based on model simulation. While the model
prediction is interesting and compelling, we excluded from the
outset (see \S1) the rapid helicity change during the flare time due
to observational limitation. Our conclusion is valid only for the
long term variation of helicity. Yokoyama et al. (2003) have found
that flares tend to occur after reversal of helicity injection rate
changed its sign. Although this is occasionally seen in our samples,
(i.e., in the case of AR 10030 and AR 10720) as well, it is not
always the case and we are unsure whether this is a necessary
condition for the flares. More often than not, the helicity either
remains constant or increases in one sign when the flare occurs.

\section{Correlation with Soft X-ray Flux}

We compare the helicity change rate, $dH/dt$ and accumulation amount,
$\Delta H$ with the time integrated soft X-ray flux taken as the
proxy for the flare energy release. In addition we check the
helicity accumulation time, $\tau$, defined as the time interval of
helicity accumulation measured from $t_0$ and the first coming
flare. Prior to make such a comparison, the range of uncertainty of
each quantity needs to be known. In general it is hard to trace
all the possible uncertainties involved with each quantities in
Equation (1). Fortunately, our targeted quantity given by Equation (2)
involves integration in space and time and the uncertainty in
each measured quantity is not propagating,
but rather may cancel out in the process of spatial and time
integration if it is random in nature. We thus focus the uncertainty
estimate only on the linear approximation of the helicity variation
that we are after. We first find out the best-fit linear function to
the points $\Delta H(t_i)$ lying in phase I (i.e., $t_0\leq t_i \leq
t_0+\tau$) in the form of $F(t)=a(t-t_0)+F(t_0)$. Next we calculate the
standard deviation, $\sigma$, of the scatter points with respect to this
linear function, and plot two additional lines corresponding to the
$\pm\sigma$ levels of the scatter points. Finally, we read the $y$-axis offsets
of these two lines to determine $\sigma_{\Delta H}$ and the x-axis offsets
to determine $\sigma_\tau$, respectively. In addition, we calculate the
uncertainty of the slope $a$ itself in the form
of $(\Delta H-\sigma_{\Delta H})/(\tau+\sigma_\tau) \leq a \leq(\Delta
H+\sigma_{\Delta H})/(\tau-\sigma_\tau)$. The center value of $a$ here
is taken as the average helicity change rate in the rest of this paper.
Therefore, by the average helicity change rate, we do not mean the
average of the quantities given in Equation (1), but we refer to the
best fit slope to the helicity variation (eq. [2]) in phase I. The
uncertainties shown in Figure 4 and Table 1 are those associated
with our linear function fit only.

In Figure 4, we plot, as symbols, the helicity parameters against
the GOES soft X-ray fluxes integrated over the flaring time ($F_X$,
hereafter). Each symbol is identified with the event ID number in
the figure together with uncertainty range represented by the bar
(see also Table I). The solid lines show the least-squares linear
fits to the data points. The correlation coefficients (CCs) of the
linear fits are also given in each panel. Figure 4$a$ shows that
there is a fairly good correlation (CC=0.86) between the helicity
change rate and $F_X$. The amount of helicity accumulation also
shows a modest correlation with $F_X$ (CC=0.68) as shown in Figure
4$b$, although not as good as for the helicity change rate. On the
other hand, the correlation between helicity accumulation time
$\tau$ and the soft X-ray flux is very poor with a weak tendency
that the longer accumulation time $\tau$, the weaker soft X-ray flux
$F_X$ (Fig. 4$c$).

We initially expected, on a general basis, that the helicity change
$\Delta H$ would strongly correlate with $F_X$. It is therefore
puzzling why the helicity change rate $dH/dt$ shows even a better
correlation with $F_X$ in Figure 4. As a possibility, we considered
that $\tau$ may be a factor in complicating the relationship between
$\Delta H$ and $F_X$. A intriguing idea is that the magnetic energy
decays much faster than the magnetic helicity in the presence of
magnetic diffusion (Berger 1999). We thus compare $\Delta H$ with
$\tau$ in Figure 4d, which unfortunately shows no obvious correlation
between them. The small number of events used in this study is another
restriction for finding a trend here. With the present result alone,
it is fair to presume that the weaker correlation between $F_X$ and
$\Delta H$ may arise from our inaccurate determination of the helicity
accumulation amount due to unknown initial time of helicity build-up.

\section{SUMMARY}

We have investigated the variation of magnetic helicity over a time
span of several days around the times of 11 X-class flares which
occurred in seven active regions using MDI magnetograms to find the
following results.

First, a substantial amount of helicity accumulation is found before
the flare in all the events. The helicity increases at a nearly
constant rate, (4.5--48)$\times$10$^{40}$ Mx$^2$ hr$^{-1}$, over a
period of 0.6 to a few days, resulting in total amount of helicity
accumulation in the range of (1.8--16)$\times$10$^{42}$ Mx$^2$. Such
a wide range of helicity accumulation indicates that each active
region has its own limit of helicity storage to keep a stable
magnetic structure in the corona. The finding of a monotonically
increasing phase is similar to the earlier one by Sakurai \& Hagino
(2003) that the magnetic helicity integrated over the regions
evolved slowly and did not show abrupt changes at the time of the
flares. The helicity increase over days before the flares reconfirms
the conventional idea that helicity accumulation by a certain amount
is necessary for a large flare to occur (Kusano et al. 1995; Choe \& Lee 1996).

Second, there is a strong positive correlation between the average
helicity change rate of phase I and the corresponding GOES X-ray
flux integrated over the flaring time. The amount of helicity
accumulation during phase I also correlates with the soft X-ray
flux, as expected, but the correlation is stronger with the helicity
change rate. This result probably implies that the helicity change
rate is more accurately determined than the amount of helicity
change itself as the initial time of helicity build-up is poorly
determined.

If the above correlations hold for a large number of events, we may
predict the flare strength (the integrated X-ray flux) based on the
helicity change rate. Monitoring of helicity variation in target
active regions may also aid the forecasting of flares. A warning
sign of flares can be given by the presence of a phase of monotonically
increasing helicity, as we found that all the major flares occur
after significant helicity accumulation. As a reference we have
checked helicity variation of the six active regions in non-flaring
times to find much lower helicity change rates compared with those
around the major flares. We thus conclude that the relative magnetic
helicity can be a powerful tool for predicting major flares.

\acknowledgments

The authors wish to thank the referee for valuable comments on the
manuscript. The work is supported by NSF grant ATM-0548952 and
NASA grant NNG0-6GC81G. J.L. was supported by NSF grant AST 06-07544
and NASA grant NNG0-6GE76G. G.S.C. was supported by DOE contract
DE-AC02-76-CH03073 and NASA grant NNH04AA16I.

\clearpage

\begin{deluxetable}{llcccccc}
\tabletypesize{\scriptsize}
\tablewidth{0pt}
\tablecaption{List of flares, helicity and accumulation time}
\tablehead{
\colhead{ID}  & \colhead{Flares}   & \colhead{AR}          & \colhead{Peak time}            & \colhead{$F_X\tablenotemark{a}$}           &
\colhead{$\vert$$dH/dt$\tablenotemark{b}$\vert$}           & \colhead{$\vert$$\Delta H$\tablenotemark{c}$\vert$}           &           \colhead{$\tau\tablenotemark{d}$}\\
\colhead{}  & \colhead{}   & \colhead{number}          & \colhead{$(UT)$}            & \colhead{$(10^{-1}J/m^2)$}           &
\colhead{$(10^{40}Mx^{2}hr^{-1})$}           & \colhead{$(10^{42}Mx^{2})$}           & \colhead{$(hr)$}}
\startdata
1 & X 1.3 on Oct 25, 2001 & 9672 & 15:02 & 2.3 & 6.2$\pm$0.9 & 1.8$\pm$0.1 & 29$\pm$2 \\
2 & X 3.0 on Jul 15, 2002 & 0030 & 20:08 & 1.4 & 13.3$\pm$1.5 & 1.9$\pm$0.1 & 14$\pm$1 \\
3 & X 1.5 on Mar 17, 2003 & 0314 & 19:05 & 1.3 & 6.4$\pm$1.6 & 3.9$\pm$0.5 & 63$\pm$8 \\
4 & X 1.5 on Mar 18, 2003 & 0314 & 12:08 & 1.3 & 13.3$\pm$1.4 & 2.3$\pm$0.1 & 17$\pm$1 \\
5 & X 18 on Oct 28, 2003 & 0486 & 11:10 & 20.0 & 48.4$\pm$6.3 & 14.5$\pm$0.9 & 30$\pm$2 \\
6 & X 10 on Oct 29, 2003 & 0486 & 20:49 & 9.1 & 46.8$\pm$3.0 & 15.9$\pm$0.5 & 34$\pm$1 \\
7 & X 1.2 on Feb 26, 2004 & 0564 & 02:03 & 0.75 & 4.5$\pm$0.4 & 3.1$\pm$0.1 & 70$\pm$3 \\
8 & X 2.2 on Nov 07, 2004 & 0696 & 16:06 & 2.1 & 19.8$\pm$0.7 & 10.7$\pm$0.2 & 54$\pm$1 \\
9 & X 1.3 on Jan 15, 2005 & 0720 & 00:43 & 1.3 & 22.2$\pm$4.1 & 4.2$\pm$0.4 & 19$\pm$2 \\
10 & X 2.8 on Jan 15, 2005 & 0720 & 23:00 & 6.6 & 22.5$\pm$1.2 & 4.3$\pm$0.1 & 19$\pm$1 \\
11 & X 4.1 on Jan 17, 2005 & 0720 & 09:52 & 9.1 & 40.8$\pm$5.5 & 3.2$\pm$0.2 & 8$\pm$1
\enddata
\tablenotetext{a}{Integrated GOES X-ray flux.}
\tablenotetext{b}{Average helicity change rate of phase I.}
\tablenotetext{c}{The amount of helicity accumulation during phase I.}
\tablenotetext{d}{Helicity accumulation time.}
\end{deluxetable}

\clearpage

\begin{figure}
\begin{center}
\includegraphics[scale=0.77]{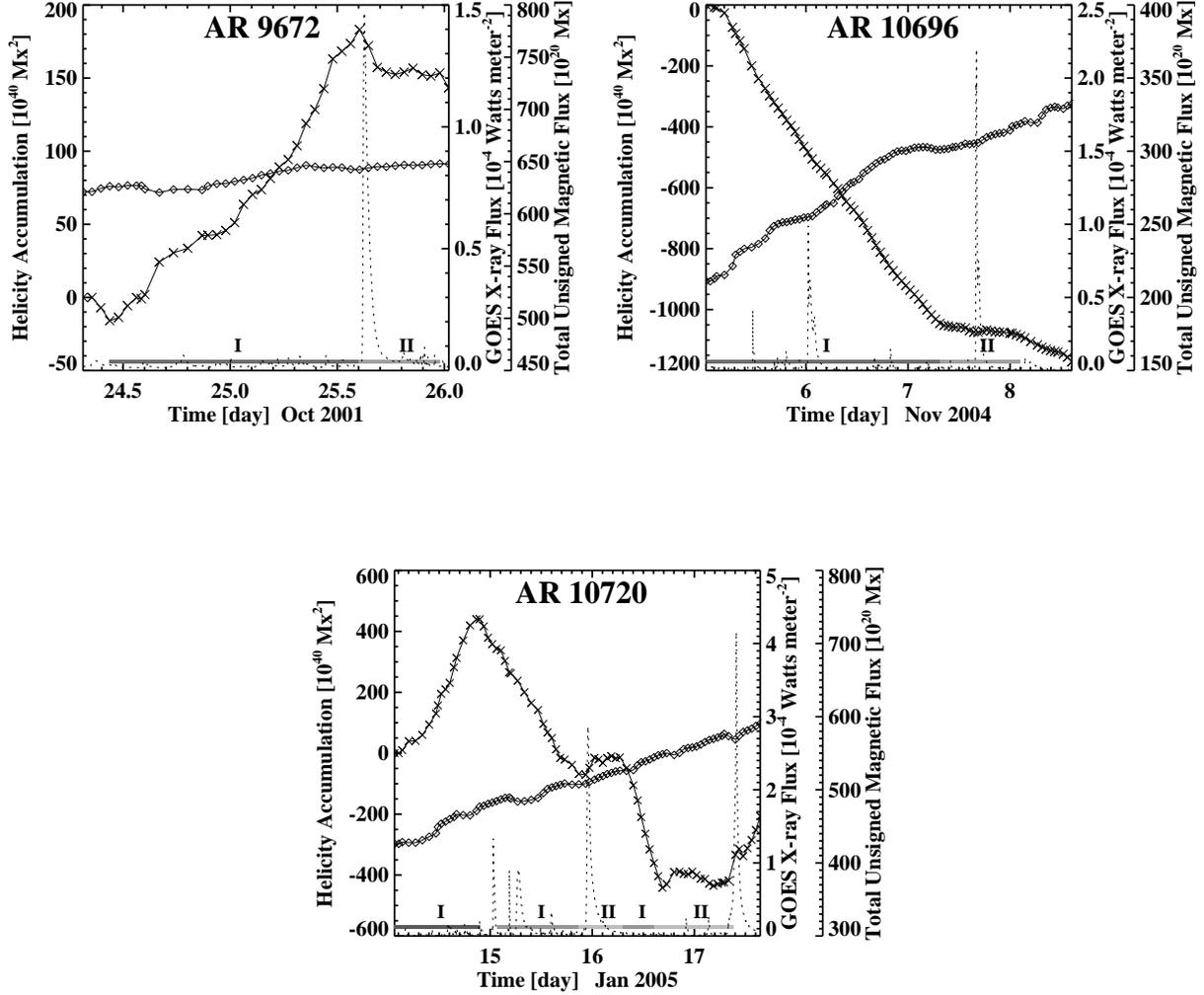}
\caption{Time variations of helicity accumulation, magnetic flux, and $GOES$ X-ray flux for three active regions. The helicity is shown as cross symbols and the magnetic flux is shown as diamonds. The $GOES$ X-ray flux is shown as the dotted lines. Phase I, the interval over which the helicity accumulation is considered, and phase II, the following phase of relatively constant helicity, are marked. \label{fig1}}
\end{center}
\end{figure}

\clearpage

\begin{figure}
\begin{center}
\includegraphics[scale=0.77]{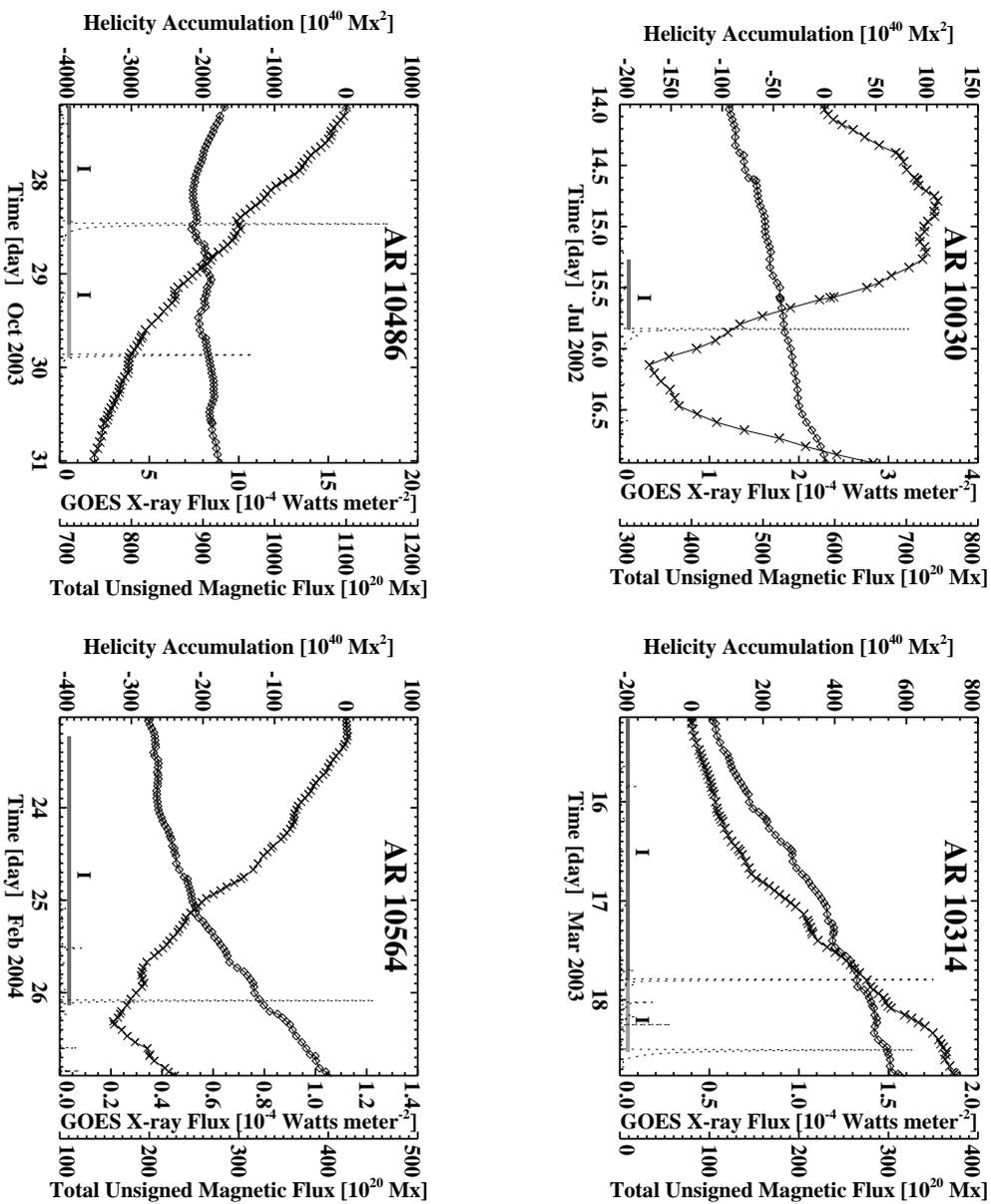}
\caption{Same as in Fig. 1, but for additional events. \label{fig2}}
\end{center}
\end{figure}

\clearpage

\begin{figure}
\begin{center}
\includegraphics[scale=0.7]{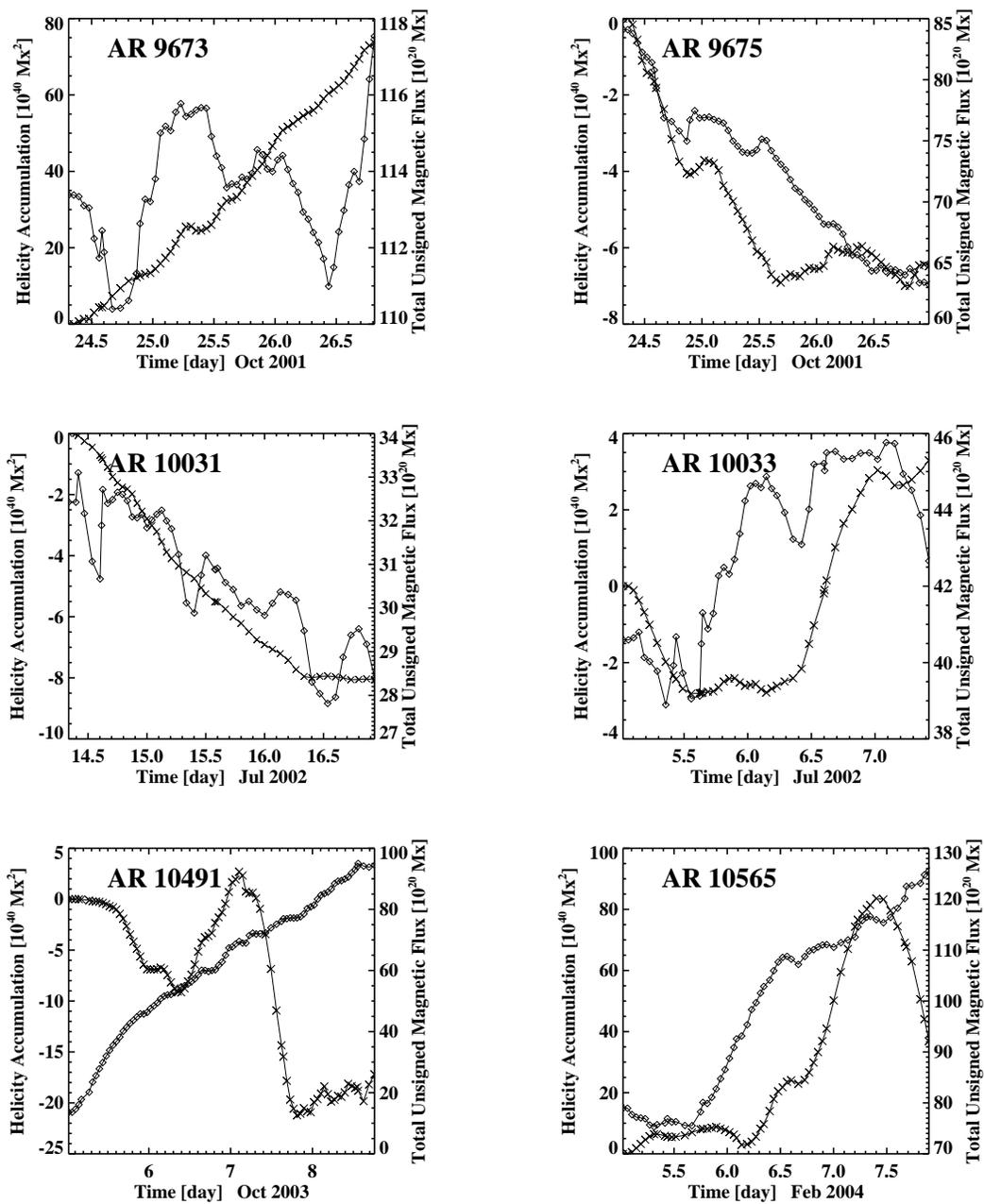}
\caption{Time variations of helicity accumulation, and magnetic flux for six nonflare active regions. The helicity is shown as crosses, and the magnetic flux is shown as diamonds.
\label{fig4}}
\end{center}
\end{figure}

\clearpage

\begin{figure}
\begin{center}
\includegraphics[scale=0.82]{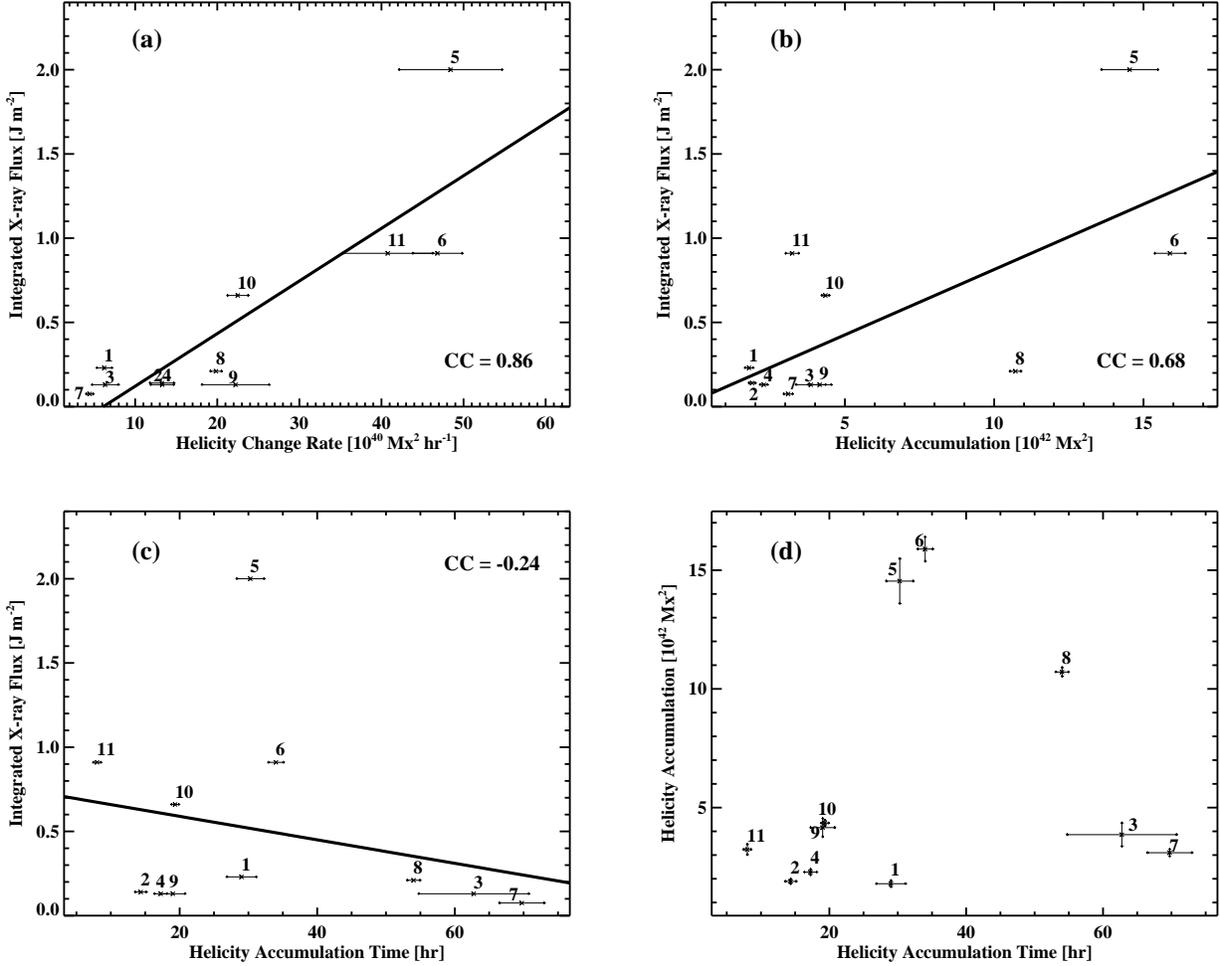}
\caption{Helicity parameters with $GOES$ X-ray flux integrated over the flaring time. Correlations of the integrated soft X-ray flux with ($a$) average helicity change rate of phase I, ($b$) the amount of helicity accumulation during phase I, and ($c$) helicity accumulation time. Correlation coefficient (CC) is specified in each panel. In ($d$), the amount of helicity accumulation is plotted as a function of the accumulation time. The uncertainties of the average helicity change rate, the amount of helicity accumulation, and the helicity accumulation time are shown as error bars in each panel.
\label{fig3}}
\end{center}
\end{figure}


\begin{thebibliography}{}
\bibitem[Alissandrakis(1981)]{ali81}  Alissandrakis, C. E.   1981, \aap, 100, 197
\bibitem[Berger \& Field(1984)]{ber_fie84} Berger, M. A., \&  Field, G. B.   1984, J. Fluid Mech., 147, 133
\bibitem[Berger(1999)]{ber99} Berger, M. A.   1999, in M. R. Brown, R. C. Canfield, and A. A. Pevtsov (eds.)  Magnetic Helicity in Space and Laboratory Plasmas, Geophys. Monogr., 111, 11
\bibitem[Brown, Canfield, and Pevtsov(1999)]{bro_can_pev99} Brown, M. R., Canfield, R. C., \& Pevtsov, A. A. (eds.)   1999, Magnetic Helicity in Space and Laboratory Plasmas, Geophys. Monogr., 111
\bibitem[Chae(2001)]{cha01} Chae, J.   2001, \apjl, 560, L95
\bibitem[Chae \& Jeong(2005)]{cha_jeo05} Chae, J., \& Jeong, H.   2005, J. Korean Astron. Soc., 38, 295
\bibitem[Choe \& Lee(1996)]{cho_lee96} Choe, G. S., \& Lee, L. C.   1996, \apj, 472, 372
\bibitem[D\'emoulin \& Berger(2003)]{dem_ber03} D\'emoulin, P., \& Berger, M. A.   2003, \solphys, 215, 203
\bibitem[Hartkorn \& Wang(2004)]{har_wan04} Hartkorn, K., \& Wang, H.   2004, \solphys, 225, 311
\bibitem[Kusano et al.(1995)]{kus95}  Kusano, K., Suzuki, Y., \& Nishikawa, K.   1995, \apj, 441, 942
\bibitem[Kusano et al.(2002)]{kus02}  Kusano, K., Maeshiro, T., Yokoyama, T., \& Sakurai, T.   2002, \apj, 577, 501
\bibitem[Kusano et al.(2003)]{kus03}  Kusano, K., Yokoyama, T., Maeshiro, T., \& Sakurai, T.   2003, Adv. Space Res., 32, 1931
\bibitem[Kusano et al.(2004)]{kus04}  Kusano, K., Maeshiro, T., Yokoyama, T., \& Sakurai, T. 2004, ApJ, 610, 537
\bibitem[Longcope et al.(2007)]{lon07}  Longcope, D. W., Ravindra, B., \& Barnes, G.   2007, \apj, 668, 571
\bibitem[Moon et al.(2002a)]{moo02a}  Moon, Y.-J., Chae, J., Wang, H., \& Park, Y. D.   2002a, \apj, 580, 528
\bibitem[Moon et al.(2002b)]{moo02b}  Moon, Y.-J., Chae, J., Choe, G. S., Wang, H., Park, Y. D., Yun, H. S., Yurchyshyn, V., \& Goode, P. R.   2002b, \apj, 574, 1066
\bibitem[November \& Simon(1988)]{nov_sim88}  November, L. J., \& Simon, G. W.   1988, \apj, 333, 427
\bibitem[Parker(1955)]{par55}  Parker, E. N.   1955, \apj, 122, 293
\bibitem[Phillips et al.(2005)]{phi05}  Phillips, A. D., MacNeice, P. J., \& Antiochos, S. K.   2005, \apj, 624, L129
\bibitem[Rust(2001)]{rus01}  Rust, D. M. 2001, J. Geophys. Res., 106, 25075
\bibitem[Sakurai \& Hagino(2003)]{sak_hag03}  Sakurai, T., \& Hagino, M.   2003, Adv. Space Res., 32, 1943
\bibitem[Scherrer et al.(1995)]{sch95}  Scherrer, P. H., et al.   1995, \solphys, 162, 129
\bibitem[Yokoyama et al.(2003)]{yok03}  Yokoyama, T., Kusano, K., Maeshiro, T., \& Sakurai, T.   2003, Adv. Space Res., 32, 1949

\end{thebibliography}
\end{document}